\documentclass[12pt]{iopart}
\usepackage{iopams}
\usepackage{graphicx}

\usepackage[english]{babel}
\usepackage[T1]{fontenc}
\usepackage[utf8]{inputenc}

\graphicspath{{./img/}}

\usepackage{xcolor}

\newcommand{\Ec}{\mathcal{E}}
\newcommand{\nn}{\textbf{\scriptsize n}}
\newcommand{\kk}{\textbf{\scriptsize k}}

\begin{document}

\title{Relativistic effects in ionization of heavy hydrogen-like ions by short X-ray pulses}
\author{O~Novak%$^*$
, M~Diachenko and R~Kholodov}
\address{Institute of applied physics, National academy of sciences of Ukraine, Petropavlivska street, 58, 40000 Sumy, Ukraine}
% \address{$^*$Author to whom any correspondence should be addressed.}
\ead{novak-o-p@ukr.net}
\date{2022}

\begin{abstract}
The probability of photoionization of a heavy hydrogen like ion is calculated within the relativistic perturbation 
theory. 
The Coulomb potential of a heavy nucleus is taken into account accurately by usage of the corresponding solutions of the Dirac equation.
The obtained probability is compared with both nonrelativistic limit and with nonperturbative results based on the numerical solution of the time-dependent Dirac equation.
Suppression of ionization is observed in the secondary ionization channel at photon frequency of about $\hbar\omega \approx mc^2$.
\end{abstract}

\maketitle

%%%%%%%%%%%%%%%%%%%%%%%%%%%%%%%%%%%%%%%%%%%%%%%%%%%%%%%%%%%%%%%%%%%%%%%%%%%%%%%
\section{Introduction}
\label{sec:intro}

Long-standing interest in laser-matter interaction is motivated by significant progress in laser technology~\cite{Reiss92, Salamin06, DiPiazza12, Milosevic06}.
Modern laser facilities are capable to reach field strength exceeding the strength of Coulomb field binding inner electrons in most atoms.
The new generation of laser facilities being constructed is expected to reach and even to exceed the critical Schwinger field and allow to observe exotic phenomena like electromagnetic vacuum decay.
Increasing in pulse power is tightly connected with shortened pulse duration. 
Presently, laser pulses as short as few or even a single optical cycle are available~\cite{Brabec00, Baltuska03, Lopez05}.
In addition, the higher photon energies are envisaged with free-electron laser projects.
In particular, we mention projects Extreme Light Infrastructure~\cite{ELI}, The European XFEL~\cite{XFEL} and Facility for Antiproton and Ion Research~\cite{FAIR, Gumberidze05, Stoehlker05, Gumberidze09}.
These new possibilities calls for the fully relativistic treatment of atomic processes, based on the relativistic Dirac equation.

The present study has been partially motivated by the proposal of Gamma Factory facility, which is currently discussed at CERN~\cite{Budker20}.
The Gamma Factory is based on a high-energy storage ring with circulating beam of highly charged ions.
An intense beam of light from a laser facility would be directed head-on to the ion beam.
In the rest frame of the ion beam, the photon energy is boosted by a factor of $2\gamma$, enabling excitation of the inner electrons of ions.
The excited ions, in turn, emit secondary photons, which in the laboratory frame are emitted predominantly in the propagation direction of the ion beam.
The Gamma Factory facility is expected to offer a source of hard photons with energies up to 400~keV and photon flux which is many orders of magnitude greater than currently available at modern gamma sources.

The inner electrons of heavy ions experience Coulomb field that comprises a noticeable fraction of the critical QED field.
This offers unique ways to explore various fundamental problems in QED.
However, transition frequencies in heavy ions lays in X-ray domain and can not be reached by modern lasers.
Therefore, various collisional processes are used to induce transitions of inner electrons of heavy ions in majority of experiments.
The Gamma Factory will open the possibility of direct spectroscopy of heavy ions.

Photoionization is one of the key processes in laser-matter interaction and it has been studied in a large number of works (see Refs.~\cite{Reiss92, Salamin06, DiPiazza12, Milosevic06, DimitrovskiPhD} for a review).
In Ref.~\cite{Pindzola10, Pindzola12}, single- and two-photon photoionization of hydrogen-like ions has been studied in dipole approximation and with account of full electromagnetic potential.
Multiphoton photoionization of hydrogen-like systems is considered in Ref.~\cite{Vanne12} within the dipole approximation using numerical approach. 
In Ref.~\cite{Selsto09}, above-threshold ionization of a hydrogen ion by a UV pulse has been studied within an approach based on the numerical solution of the Dirac equation.
In Ref.~\cite{Simonsen16}, envelope induced ionization dynamics beyond the dipole approximation has been considered.
In Ref.~\cite{Dondera12, Bachau13},  photoionization of hydrogen and hydrogen-like ions is studied with account of a relativistic correction to the Schr\"odinger equation.

Purely numerical approaches are limited by the available computational resources, while the dependence of the ionization probability on the physical parameters often can be hard to infer from the numerical results.
On the other hand, most of analytical approaches are based on the original work of Keldysh and later of F.~Faisal and H.~Reiss~\cite{Keldysh65, Faisal73, Reiss80}. 
In this approach called strong field approximation, atomic potential is treated as a perturbation to strong variable electric field of an electromagnetic wave. 
In the adiabatic limit, photoionization is described as tunneling.
For example, in a recent work \cite{Eckey22} relativistic strong-field ionization of hydrogen-like atoms or ions in a constant crossed electromagnetic field is studied within the modified trong-field approximation in Göppert-Mayer gauge.

However, these approaches may not be applicable when pulse duration equals to only few optical cycles and electric field variation is not adiabatic.

In this paper we use the relativistic perturbation theory to calculate the probability of photoionization of a heavy hydrogen-like ion. 
The Coulomb potential is taken into account accurately and the laser wave is considered as a perturbation. 
This approach imply some constraints to laser field strength and intensity.
More specifically, laser field strength must be less than the typical Coulomb field experienced by a bound electron.
To ensure applicability of perturbative approach, we choose the laser field to be equal to one percent of the Coulomb field for all analytical results across the manuscript. 
Validity of analytical results are confirmed by comparison with the nonrelativistic limit as well as with the nonperturbative calculations based on numerical solution of the time-dependent Dirac equation.

The relativistic unit system is used, $\hbar = m = c = 1$, if not stated othewise.

%==============================================================================
\section{Probability of heavy ion photoionization}
\label{sec:perturbation}

In the present work we consider laser wave as a perturbation, while the Coulomb field of the nucleus is treated within the relativistic theory.
The qualitative change in behavior of atomic physics processes is expected when the electric field of the laser is of the same order of magnitude as the Coulomb electric field experienced by an electron in a ground state, $E'$.
Similarly, the characteristic measure of the laser photon energy could be conveniently chosen of the order of the corresponding binding energy.
Thus, for the field strength and the photon frequency we introduce units 
\begin{equation}
\label{sc}
	\begin{array}{l}
		\omega' = 2\mathcal{E}_{bind}, \\ \displaystyle
		E' =  \frac{3Z\alpha}{\langle r^2 \rangle},
	\end{array}
\end{equation}
where $\alpha$ is the fine structure constant, $\mathcal{E}_{bind}$ is the binding energy and $\langle r^2 \rangle$ is the mean square of the radial electron coordinate in the ground state.
Note that $\omega'$ and $E'$ coincide with atomic units when $Z=1$ and the nonrelativistic expression is used for $\Ec_{bind}$ and $\langle r^2 \rangle$.

Using the known solution to the Dirac equation for an electron in a field of a point-like nucleus it is possible to find the explicit expressions for $E'$ and $\omega'$ in the form
\begin{equation}
\label{rsc}
  \begin{array}{l}
	  \omega' = 2(1-\mathcal{E}_{0}), \\ \displaystyle
	  E' = \frac{\alpha^{5/2} 12 Z^3 }{ (2 \mathcal{E}_{0}+1) (2\mathcal{E}_{0}+2)}.
  \end{array}
\end{equation}
where $\mathcal{E}_0$ is the ground energy,
\begin{equation}
  \mathcal{E}_0 = \sqrt{1 - (\alpha Z)^2}
\end{equation}
 
We adopt the dipole approximation, so that the magnetic field vector of the laser is zero and the electric field vector of the laser depends only on time, $E(t)$.
The effects accounted for within this approximation dominate the physics for a wide range of intensities and frequencies. 
In the strong field domain, the dipole approximation is justified, if the electron displacement caused by an electromagnetic wave is less than wavelength and the typical electron orbit size, $\left<r^2\right>$~\cite{Reiss92}.
This conditions can be expressed in terms of ponderomotive potential $U_p = \alpha E^2/4\omega^2$ as
\begin{equation}
\label{dipoleConditions}
  U_p \ll 1, \qquad  \frac{\alpha Z U_p}{\omega} \ll 1
\end{equation}
In the present work we consider the case of low field strength compared to the critical field $E'$, and laser frequency comparable to $\omega'$.
It is easy to see that these values satisfy the inequalities (\ref{dipoleConditions}). 
For example, for the case of photoionization of a hydrogenlike Radon ion by a laser pulse with $E=10^{-2}E'$ and $\omega \approx \omega'$ we obtain $U_p \sim 10^{-5}$ and $ \alpha Z U_p/\omega \sim  10^{-5}$.
Typical laser intensity for which the dipole approximation breaks down due to strong field effects can be estimated as $I \sim 10^{29}$~W/cm$^2$ in this case.

In case of high frequencies, however, another requirement is relevant. 
Specifically, the size of electron orbital $R_Z$ should be less than the laser wavelength $\lambda$ [Vanne, Phys. Rev. A, 2012].
Taking a Radon ion as an example and typical frequency value of $\omega \approx \omega'$,  we obtain
$R_Z/\lambda \approx 0.1$. 
We consider this estimation as expected size of the neglected non-dipole contribution to the presented results.

To find the photoionization probability, we start with the time-dependent 
Dirac equation for combined potential of a pointlike nucleus and a laser wave,
\begin{equation}
\label{tdde}
	i\frac{\partial\Psi}{\partial t} = \hat{\mathcal{H}} \Psi.
\end{equation}
Within the dipole approximation the electric field of the laser pulse can be described by scalar potential written as 
\begin{equation}
\label{phi}
	\phi = - E(t) z,
\end{equation}
where $E(t)$ is time-dependent electric field of the laser pulse. 
The wave polarization is chosen along $Oz$ axis, and the total momentum projection $\mu$ on this axis is conserved. 
Thus, the Hamiltonian in Eq.~(\ref{tdde}) reads
\begin{equation}
\label{hamiltonian}
	\hat{\mathcal{H}} = \hat{\mathcal{H}}_0 + e \phi,
\end{equation}
where $e = -\sqrt{\alpha}$ is the electron charge, 
and $\hat{\mathcal{H}}_0$ is the nonperturbed Hamiltonian for the Coulomb potential of a point nucleus, 
\begin{equation}
\label{H0}
	\hat{\mathcal{H}}_0 = \hat{\vec{\alpha}} \vec p + \hat\beta - \frac{Z\alpha}{r}.
\end{equation}
Here, $\hat{\vec{\alpha}}$ and $\hat\beta$ are the Dirac matrices and $Z$ is the nucleus charge.

To solve Eq.~(\ref{tdde}), we use the ansatz
\begin{equation}
\label{psi}
	\Psi = \sum a_\nn(t) \Phi_\nn e^{-i\mathcal{E}_\nn t},
\end{equation}
where the bold-faced indices denote sets of quantum numbers.
The wave functions $\Phi_n$ satisfy the equation
\begin{equation}
	\hat{\mathcal{H}}_0 \Phi_\nn = \mathcal{E}_\nn \Phi_\nn.
\end{equation}
Inserting (\ref{psi}) to Eq.~(\ref{tdde}), one obtains the following set of
coupled differential equation for the amplitudes $a_\nn(t)$,
\begin{equation}
\label{cce}
	i \dot{a}_\kk (t) = \sum a_\nn(t) e^{i(\mathcal{E}_\kk - \mathcal{E}_\nn)t} V_{\kk\nn},
\end{equation}
with matrix element $V_{kn}$ defined as
\begin{equation}
\label{vkn}
	V_{\kk\nn} = \int d\:^3r \: \Phi^+_\kk e\phi \: \Phi_\nn.
\end{equation}

We restrict our study to transitions from the ground state to the positive continuum. 
Thus, with account of Eq.~(\ref{phi}), the corresponding matrix element (\ref{vkn}) can be transformed to
\begin{equation}
	V_{\varepsilon} = \sqrt{\alpha}E(t) z_{\varkappa}(\mathcal{E}) ,
\end{equation}
\begin{equation}
\label{defze}
	z_{\varkappa}(\mathcal{E})  = \int d^3r \: \Phi^+_{\varepsilon\varkappa} z \Phi_{0},
\end{equation}
where the subscript 0 denotes the ground state, $\mathcal{E}$ is the final electron energy and $\varkappa$ is the spin-orbit quantum number determined by the angular momentum $j$ and the parity of the state $(-1)^l$, 
\begin{equation}
  \varkappa = (-1)^{j+l+1/2} (j + 1/2), 
\end{equation}
so that
\begin{eqnarray}
  l=\cases{\varkappa,& $\varkappa > 0$, \\ -\varkappa - 1,&  $\varkappa < 0$, \\}
  \\
  j = |\varkappa| - \frac{1}{2}.
\end{eqnarray}

Note that $V_\varepsilon$ factorizes to spatial and time-dependent parts.
Finally, taking $a_{0} \approx 1$ for the ground state and $a_\nn \approx 0$ for all others,
the amplitudes on the right-hand side of Eq.~(\ref{cce}) are approximated by constants, 
and one obtains a simple expression 
\begin{equation}
\label{ae}
	a_{\varkappa}(\mathcal{E}) \approx 
	-i\sqrt{\alpha} \: z_{\varkappa}(\mathcal{E}) 
	\int\limits_{-\infty}^{+\infty} E(t) e^{i(\mathcal{E} - \mathcal{E}_{0})t} \: dt,
\end{equation}
where $\mathcal{E}_{0} = \sqrt{1-(Z\alpha)^2}$ is the ground state energy.
Note, that the quantity $a_{\varkappa}(\mathcal{E})$ can be interpreted as the amplitude of ionization probability if the condition $\phi(t=\pm\infty) = 0$ is fulfilled, otherwise it contains a term describing wave functions deformation in a constant field.

To calculate the matrix element $z_{\varkappa}(\mathcal{E})$ we use the known
eigenfunctions of the nonperturbed Hamiltonian (\ref{H0}),
\begin{equation}
\label{Phi}
	\Phi^{\mu}_{n\varkappa} = \frac{1}{r}
	\left(\begin{array}{c}
		G_{n\varkappa}(r) \chi^{\mu}_{\varkappa}\\
		iF_{n\varkappa}(r)\chi^{\mu}_{-\varkappa}\\
	\end{array}\right).
\end{equation}
Here, $G_{n\varkappa}(r)$ and $F_{n\varkappa}(r)$ are the real radial functions, 
$\chi^{\mu}_{\varkappa}$ are the spherical spinors, $n$ is the principal quantum number and
$\mu$ is the projection of the total angular momentum to $Oz$ axis.

Wave function  $\Phi_{n\varkappa}^\mu$ describes an electronic state with
defined values of electron energy $\mathcal{E}_n$, total angular momentum $j$, 
its projection $\mu$ and parity $(-1)^l$.
For an electron in the ground state the quantum numbers are
\begin{equation}
\label{ini}
	\varkappa_0 = -1, \qquad
	l_0 = 0, \qquad
	j_0 = 1/2. 
\end{equation}
Due to the problem symmetry, the angular momentum projection conserves. 
Therefore, without loss of generality we can set $\mu=1/2$ for the both initial and final states.

With account of Eq.~(\ref{Phi}) and noting that $z = r\cos\theta$, the matrix
element (\ref{defze}) takes on the form
\begin{equation}
\label{ze}
	z_{\varkappa}(\mathcal{E})  = 
	g_{\varepsilon}\Omega_{\varkappa,\varkappa_0} + f_{\varepsilon}\Omega_{-\varkappa, -\varkappa_0} ,
\end{equation}
where 
\begin{equation}
\label{gfkn}
	\begin{array}{l}
		g_{\varepsilon} = \int\limits_0^\infty   G_{\varepsilon\varkappa} G_{0} r \:dr, \\
		f_{\varepsilon} = \int\limits_0^\infty   F_{\varepsilon\varkappa} F_{0} r \:dr, 
	\end{array}
\end{equation}
\begin{equation}
	\Omega_{\varkappa,\varkappa'} = \int (\chi^{\mu}_{\varkappa})^+ \chi^{\mu}_{\varkappa'} \cos\theta \:d\Omega.
\end{equation}

Angular factors $\Omega_{\varkappa,\varkappa'}$ can be readily calculated using
angular momentum algebra~\cite{Varshalovich}, 
\begin{eqnarray}
\label{angular}
	\Omega_{\varkappa,\varkappa'} = (-1)^\varphi\Pi_{jj'll'} 
	&
  \left(\begin{array}{ccc}
		   j & 1 &   j'\\
		-\mu & 0 & \mu
	\end{array}\right)
% 	\times \\ \nonumber &\times
  \left(\begin{array}{ccc}
		l' & 1 & l \\
		0 & 0 & 0 
	\end{array}\right)
  \left\{\begin{array}{ccc}
		l' & \frac12 & j' \\
		j & 1 & l
	\end{array}\right\},
\end{eqnarray}
where the matrices in round and curled brackets denote 3j and 6j symbols respectively and
\begin{eqnarray}
  \varphi = {j+j'+l-l'-\mu+\frac12}, \\
	\Pi_{a,b,...} = \sqrt{(2a+1)(2b+1)...}
\end{eqnarray}
Taking into account selection rules for 3j and 6j symbols, and Eq.~(\ref{ini}), 
one may conclude that only two sets of final values of quantum numbers are allowed, 
\begin{eqnarray}
\label{chan1}
  l = 1, \quad j = 3/2, \quad \varkappa = -2; \\
\label{chan2}
  l = 1, \quad j = 1/2, \quad \varkappa = +1.
\end{eqnarray}
The corresponding angular coefficients can be obtained by direct calculation,
\begin{equation}
\label{o-kp}
	\begin{array}{l}
		\displaystyle
		\Omega_{-2,-1} = \Omega_{+2,+1} = \frac{\sqrt{2}}{3}, \\
		\displaystyle
		\Omega_{+1,-1} = \Omega_{-1,+1} = - \frac{1}{3}.
	\end{array}
\end{equation}

To calculate the radial integrals $g_{\varepsilon}$ and $f_{\varepsilon}$ given by 
Eqs.~(\ref{gfkn}), we use the well known wave functions for an electron in a 
Coulomb field~\cite{QEDSF, LandauIV}.
If an electron occupies ground state, these functions read
\begin{equation}
\label{gfdis}
	\begin{array}{cr}
		G_{0} &=  \hphantom{-}\sqrt{1 + \mathcal{E}_{0}} N_0 (2\zeta r)^{\gamma_0} e^{-\zeta r}, \\
		F_{0} &= -\sqrt{1 - \mathcal{E}_{0}} N_0 (2\zeta r)^{\gamma_0} e^{-\zeta r},
	\end{array}
\end{equation}
where 
\begin{eqnarray}
	\zeta &= Z\alpha, \\
	\gamma_0 &= \sqrt{1-\zeta^2}, \\
	N_0 &= \sqrt{\frac{\zeta }{\Gamma(2\gamma_0+1)}}.
\end{eqnarray}

The functions given by Eqs.~(\ref{gfdis}) are normalized according to
\begin{equation}
	\int\limits_0^\infty \left( G_{0}^2(r) + F_{0}^2(r)\right) \:dr = 1.
\end{equation}

The final state of an electron belongs to the positive energy continuum.
The corresponding wave functions have the form~\cite{QEDSF, LandauIV}
\begin{equation}
\label{gfcont}
  \begin{array}{l}
	  G_{\varepsilon\varkappa} = +\sqrt{\mathcal{E} + 1} \:N (2pr)^\gamma \mbox{Re}(\xi), \\
	  F_{\varepsilon\varkappa} = -\sqrt{\mathcal{E} - 1} \:N (2pr)^\gamma \mbox{Im}(\xi),
  \end{array}
\end{equation}
where
\begin{equation}
	\xi = e^{-ipr+i\eta} (\gamma +iy) F(\gamma+1+iy, 2\gamma+1, 2ipr),
\end{equation}
\begin{equation}
	e^{-2i\eta} = \frac{\gamma + iy}{-\varkappa + iy/\mathcal{E}},
\end{equation}
\begin{equation}
	\gamma = \sqrt{\varkappa^2 - \zeta^2},
\end{equation}
\begin{equation}
	y = Z\alpha \mathcal{E}/p,
\end{equation}
\begin{equation}
	\mathcal{E} = \sqrt{1+p^2}.
\end{equation}
We assume that the continuum wave functions are normalized according to 
\begin{equation}
	\int \Phi_{\varepsilon\varkappa}^+\Phi_{\varepsilon'\varkappa'} d^3r = 
	\delta_{\varkappa\varkappa'}\delta(\mathcal{E} - \mathcal{E}').
\end{equation}
In this case, the normalizing constant $N$ reads
\begin{equation}
	N = \frac{\left|\Gamma(\gamma+iy)\right| e^{\pi y/2}}{\sqrt{\pi p} \Gamma(2\gamma+1)}.
\end{equation}

Substitution of (\ref{gfdis}) and (\ref{gfcont}) into the matrix elements (\ref{gfkn}) 
results in integrals over the radial coordinate $r$ of the form
\begin{equation}
\label{Jdef}
	J_{\alpha\beta}^{\nu} = \int\limits_0^\infty e^{-\lambda z} z^\nu F(\alpha, \beta, kz)\:dz.
\end{equation}
This integral can be calculated explicitly and expressed via the hypergeometric 
Gauss function~${}_2F_1$:
\begin{equation}
\label{Jres}
	J_{\alpha\beta}^{\nu} = \Gamma(\nu+1)\lambda^{-\nu-1}  {}_2F_1(\alpha, \nu+1; \beta, \frac{k}{\lambda}).
\end{equation}
Taking into account (\ref{Jres}) and performing simple transformations one
obtains the integrals (\ref{gfkn}),
\begin{equation}
\label{gfeps}
	\begin{array}{cr}
		g_{\varepsilon} &= \sqrt{(\mathcal{E}+1)(1+\mathcal{E}_{0})}\: A\mbox{ Re}\:B ,\\ 
		f_{\varepsilon} &= \sqrt{(\mathcal{E}-1)(1-\mathcal{E}_{0})}\: A\mbox{ Im}\:B ,\\
	\end{array}
\end{equation}
where
\begin{equation}
\label{Adef}
	A = N_0 N (2\zeta)^{\gamma_0} (2p)^{\gamma} \Gamma(\gamma+\gamma_0+2),
\end{equation}
\begin{eqnarray}
\label{Bdef}
	B = 
	(\gamma+iy)e^{i\eta} (\zeta + ip)^{-(\gamma+\gamma_0+2)} 
	\times \\\nonumber \times
	{}_2F_1\left(1+\gamma+iy, \gamma+\gamma_0+2; 2\gamma+1; \frac{2ip}{\zeta + ip}\right) .
\end{eqnarray}
With account of (\ref{o-kp}) the matrix element (\ref{ze}) can be written as
\begin{equation}
\label{zres}
	z_\varkappa(\mathcal{E}) = \Omega_{\varkappa,-1} (f_\varepsilon + g_\varepsilon).
\end{equation}

Probability of ionization from the ground state into an energy interval 
$d\mathcal{E}$  is given by 
\begin{equation}
	dw_{\varkappa}(\mathcal{E}) = 
	\left| a_{\varkappa}(\mathcal{E}) \right|^2 d\mathcal{E}.
\end{equation}
Finally, with account of (\ref{ae}) and (\ref{zres}), the differential probability of photoionization takes on the form
\begin{equation}
\label{dwke}
	\frac{dw_\varkappa}{d\mathcal{E}} = \alpha \Omega^2_{\varkappa,-1}
	(g_\varepsilon + f_\varepsilon)^2
	\left|
		\int\limits_{-\infty}^\infty E(t) e^{i(\mathcal{E}-\mathcal{E}_{0})t} \:dt
	\right|^2 ,
\end{equation}
where the values of $\Omega_{\varkappa,-1}$ are given by Eqs.~(\ref{o-kp}).
The total ionization probability $w_{\varkappa}$ can be obtained by integration 
of Eq.~(\ref{dwke}) over the electron energy.

%==============================================================================
\textbf{Nonrelativistic case.} 
Calculations in the nonrelativistic case are similar to those performed above.
However, the conservation laws for spin and orbital momentum are obeyed separately. 
As a consequence, there is a single allowed channel of photoionization from 
the ground state with orbital momentum change
\begin{equation}
	l_0 = 0 \quad\rightarrow\quad l=1,
\end{equation}
while its projection $m$ and spin component $s_z$ conserve.
Moreover, the integral $J_{\alpha\beta}^{\nu}$ defined by (\ref{Jdef}) 
reduces to polynomials, therefore matrix element of $z$ can be expressed
in terms of elementary functions~\cite{LandauIII}.
The resulting nonrelativistic probability of photoionization from the 
ground state into an energy interval $d\mathcal{E}$ takes on the form
\begin{equation}
\label{dwnr}
	\frac{dw}{d\mathcal{E}} = \alpha\frac{2^8 f(p/\zeta)}{3\zeta^4 (1+(p/\zeta)^2)^5}
	\left|
	\int\limits_{-\infty}^t  E(t) e^{i\omega_{fi}t} dt
	\right|^2    ,
\end{equation}
where $\zeta = \alpha Z$ and $\omega_{fi} = (p^2+\zeta^2)/2$.
The auxiliary function $f(x)$ is defined as 
\begin{equation}
	f(x) = \frac{\exp\left[ -4 \frac{\mbox{arctg}(x)} {x} \right]}    { 1 + \exp(-\frac{2\pi}{x})}.
\end{equation}

It is known that Schr\"odinger equation for an atom interacting with 
external electric field allows scaling with respect to the nucleus charge.
For instance, the probability (\ref{dwnr}) stays the same for various $Z$ if laser frequency and field strength are 
measured in the units defined as
\begin{equation}
\label{nrsc}
	\begin{array}{l}
		\omega' = \alpha^2 Z^2, \\
		E' = \alpha^{5/2} Z^3.
	\end{array}
\end{equation}

On the other hand, the relativistic Dirac equation is not invariant with respect to such scaling transformations. 
Nevertheless, to facilitate comparison of Eq.~(\ref{dwke}) with its nonrelativistic counterpart Eq.~(\ref{dwnr}) we will use the units Eqs.~(\ref{rsc}) and Eqs.~(\ref{nrsc}) respectively.

%==============================================================================
\textbf{Numerical approach.}
To confirm the obtained perturbative analytical results, we will compare them to numerical nonperturbative calculations.
To solve the time-dependent Dirac equation numerically, we construct the time-dependent wave function in the form
\begin{equation}
\label{psi_num}
    \Psi = \sum\limits_{\nn} c_\nn(t) \varphi_\nn.
\end{equation}
This expression is similar to Eq.~(\ref{psi}) except the factors $e^{-i\Ec_\nn t}$ are included into the amplitudes $c_\nn(t)$ for performance reasons.
The basis wave functions $\varphi_\nn$ are the solutions of the Dirac equation for an electron moving in the Coulomb potential of an extended nuclei within the charged shell model,
\begin{equation}
\label{Vc_ext}
    V_C = -\frac{\alpha Z}{\max\left({r, R_N}\right)},
\end{equation}
where $R_N$ is the nucleus radius \cite{Angeli13}.
The unperturbed wave functions $\phi_n$ have been calculated using a variative approach based on works~\cite{Johnson88, Selsto09, McConnell12}.
This technique utilizes a finite basis set constructed from B-splines together with the dual kinetic balance approach to ensure absence of spurious solutions~\cite{Shabaev04}.

The laser potential is treated in a similar way as in the analytical approach.

After inserting the ansatz (\ref{psi_num}) into the Dirac equation, we obtain the set of ordinary differential equations in the form
\begin{equation}
\label{cce_num}
    i\dot{c} = \tilde{M}c,
\end{equation}
where elements of the matrix $\tilde{M}$ are defined as
\begin{equation}
    \tilde{M}_{\nn\kk} = \delta_{\nn\kk}\Ec_\kk + \sqrt{\alpha}E(t) <\varphi_\nn|r\cos\theta|\varphi_\kk>.
\end{equation}
In contrast to perturbative approach, we do not make any assumptions about magnitude of the amplitudes $c_\nn$ and solve the set (\ref{cce_num}) numerically.
We split time into small intervals $\Delta t$ and approximate the matrix $\tilde{M}$ by a constant value at each interval $\Delta t$.
In this case, the solution of Eq.~(\ref{cce}) can be written as
\begin{equation}
  \label{cevo}
  c(t+\Delta t) = e^{-i\tilde{M}\Delta t} c(t).
\end{equation}
To avoid direct computation of the matrix exponent, which may be very demanding to computational resources, we  find the  amplitudes $c(t)$ at each time interval using the highly efficient Lanczos propagation \cite{Park86}.

After having computed the amplitudes $c_\nn(t)$ at the end of an X-ray pulse, we can determine the probability of ionization into small energy interval $[\Ec, \Ec+\Delta \Ec]$ as
\begin{equation}
    \frac{dw(\Ec, \varkappa)}{d\Ec} = \frac{1}{\Delta\Ec}\sum\limits_{n:\Ec_n \in \Delta \Ec} |c_{n\varkappa}|^2.
\end{equation}

To compute the wavefunctions $\varphi_\nn$ we used 500 B-splines within a box of size $r_m \approx 300$ r.u.
For the numerical solution of Eq.~(\ref{cce_num}) a set of about 3500 wave functions $\Phi_n$ has been selected with eigenenergies within the interval $[-1.0, 2.5]$ and spin-orbit number values satisfying $|\varkappa| \leqslant 8$. 

%==============================================================================
\section{Results}

In this Section we compare the probability of photoionization given by 
Eq.~(\ref{dwke}) with nonrelativistic perturbation theory and numerical calculations.
Let us define electric field of a laser pulse in the form
\begin{equation}
\label{Et}
	\vec{E}(t) = \vec{e_z}E_0 \sin^2\left(\frac{\pi t}{\tau}\right)
	\sin(\omega t),
\end{equation}
where $\tau$ is the pulse duration. 
We will consider a short laser pulse with duration of five optical cycles, 
$\tau = 5\cdot 2\pi/\omega$.
Note that  in the relativistic case to calculate probability given by Eq.~(\ref{dwke}) we scale laser frequency and field strength according to (\ref{rsc}), while in the nonrelativistic limit we use nonrelativistic scaling  (\ref{nrsc}).

As is seen from (\ref{dwke}), within perturbative approach the maximum field strength
enters the photoionization probability in the form of a separate factor $E_0^2$.
Obviously, Eq.~(\ref{dwke}) is applicable only for small enough values of $E_0$, 
when total probability satisfies a condition $w_\varkappa \ll 1$.
Without loss of generality we assume $E_0 = 10^{-2}E'$ hereafter.
Note, that despite the fact that the laser field strength is set to be only a fraction of Coulomb field $E'$, it is still can be considered as extremely high in practical terms.
For example, the Coulomb field of a Rn ion is $E\approx 4.3\cdot 10^{17}$~V/m, so that field strength of $10^{-2}E'$ corresponds to laser intensity of $\sim 10^{24}$~W/cm$^2$.

\begin{figure*}
	\centering
	\includegraphics[width=0.45\textwidth]{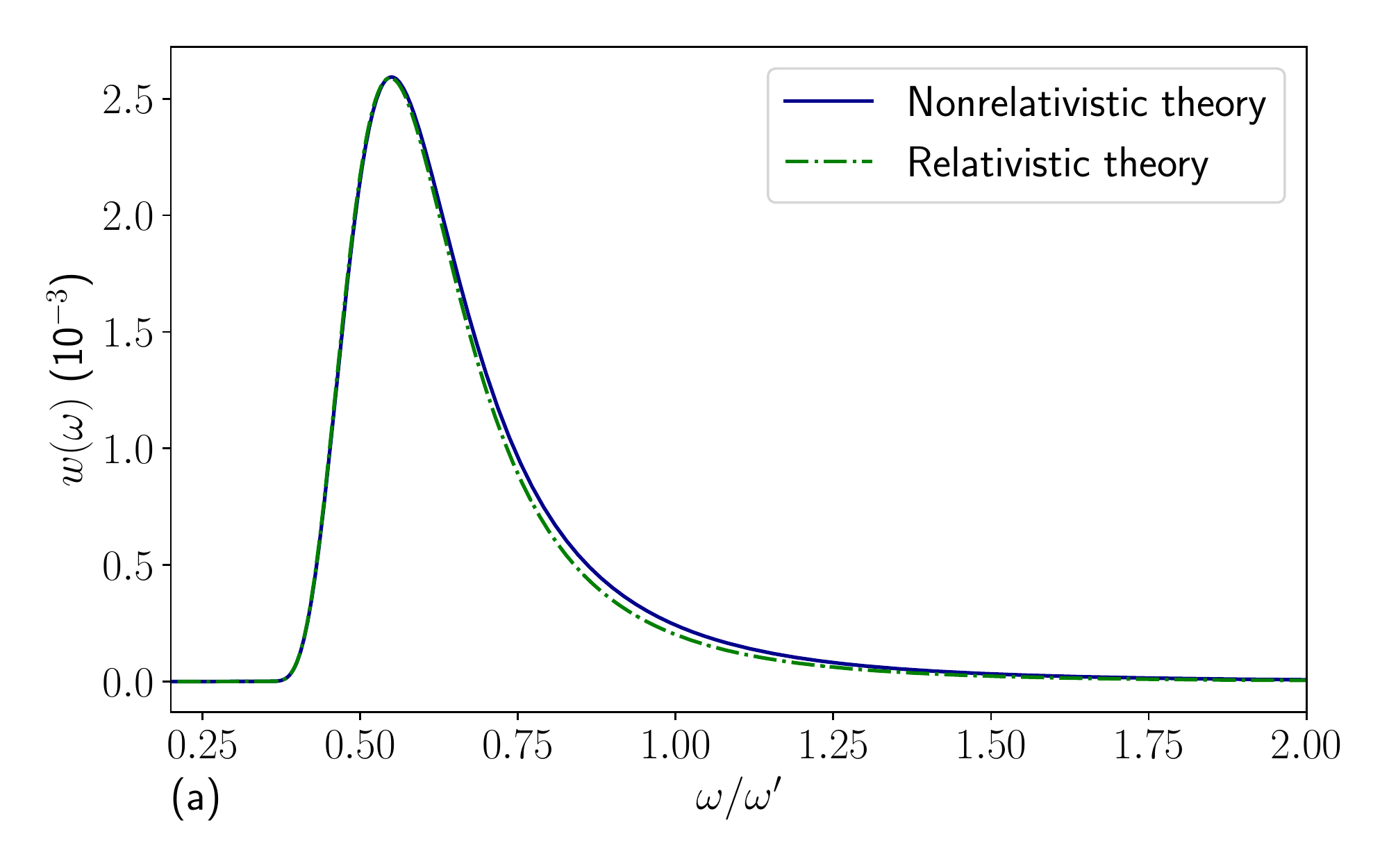}
	\includegraphics[width=0.45\textwidth]{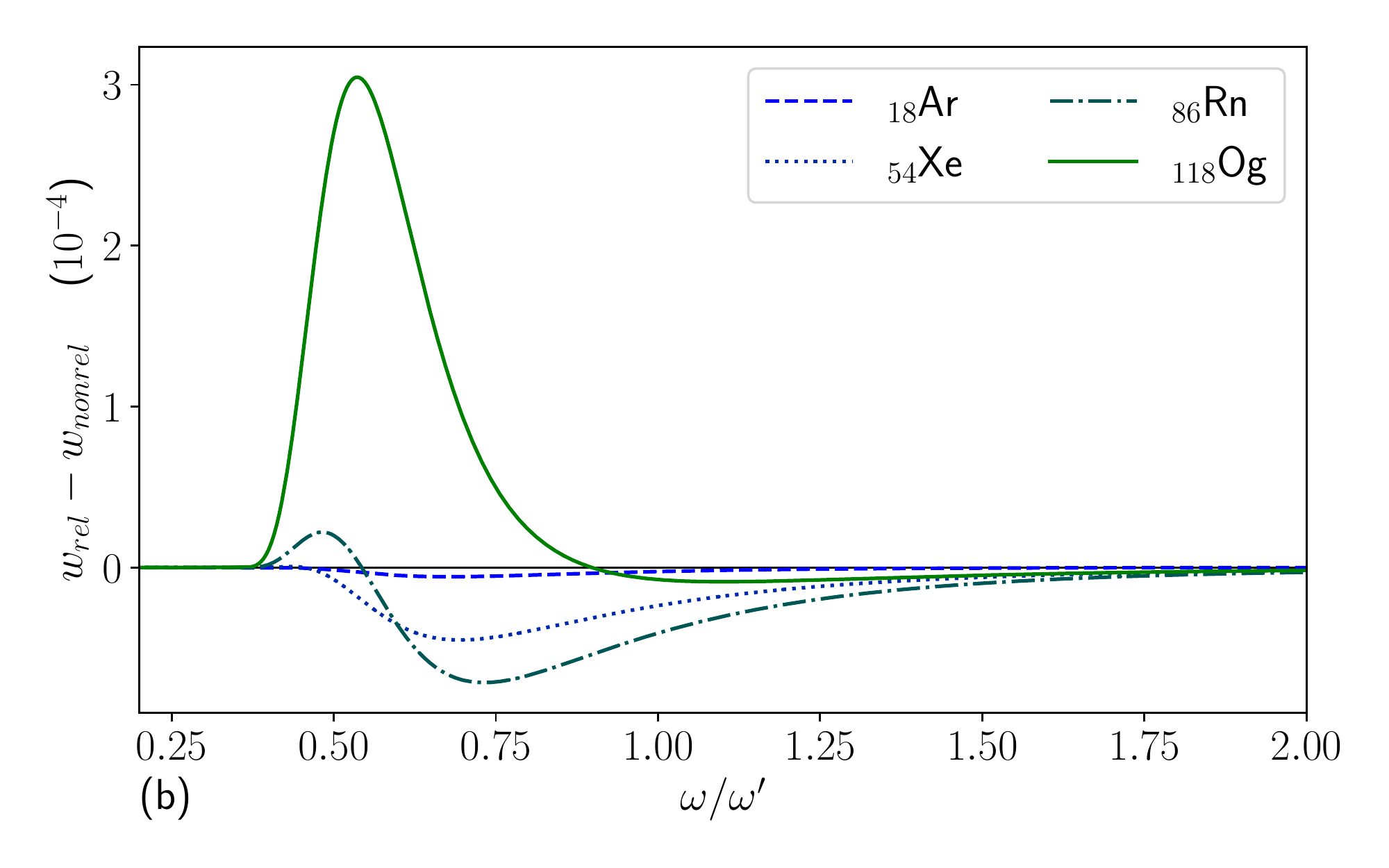}
	\caption{(a) The photoionization probability of $_{86}$Rn$^{85+}$ as a function of laser frequency.
	(b) The difference between ionization probabilities obtained within relativistic and nonrelativistic approaches for various values of charge $Z$.}
	\label{f:wdep}
\end{figure*}

Figure \ref{f:wdep} shows the dependence of the total photoionization probability on the laser frequency for the hydrogen-like Radon ion. The probability shows a maximum near the value $\omega \approx \mathcal{E}_{bind}$, with more steep decrease toward low frequency.
Note that in the nonrelativistic theory and within the dipole approximation the photoionization probability of hydrogen is the same as in the case of a nucleus with the atomic number $Z$ and with laser frequency and field simultaneously scaled as $\alpha^2Z^2$ and $\alpha^{5/2}Z^3$ respectively (in relativistic units).
In general, the relativistic effects suppress the probability compared to the scaled non-relativistic result.
However, this is mostly a consequence of more steep dependence of the binding energy, mean radial coordinate value and, hence, the Coulomb field experienced by an electron in comparison with the nonrelativistic scaling.
This effect is accounted when the scaling (\ref{sc}) is used with $\mathcal{E}_{bind}$ and $\left<r^2\right>$ given by relativistic expressions.
Similar approach is described in Ref.~\cite{Vanne12}, where an effective charge $Z'$ is introduced.
As a result, the difference in photoionization probability for various chemical elements is significantly less than than in the case of nonrelativistic scaling.
At the same time, the difference shows complex non-monotonic behavior with alternating sign.
This is true for all $Z$, as can be seen on the curve for a Rn ion in Fig~\ref{f:wdep}(b).
When $Z$ is growing,  the maximum of the curve shifts to a region with positive sign of the difference.

\begin{figure*}
	\centering
	\includegraphics[width=0.45\textwidth]{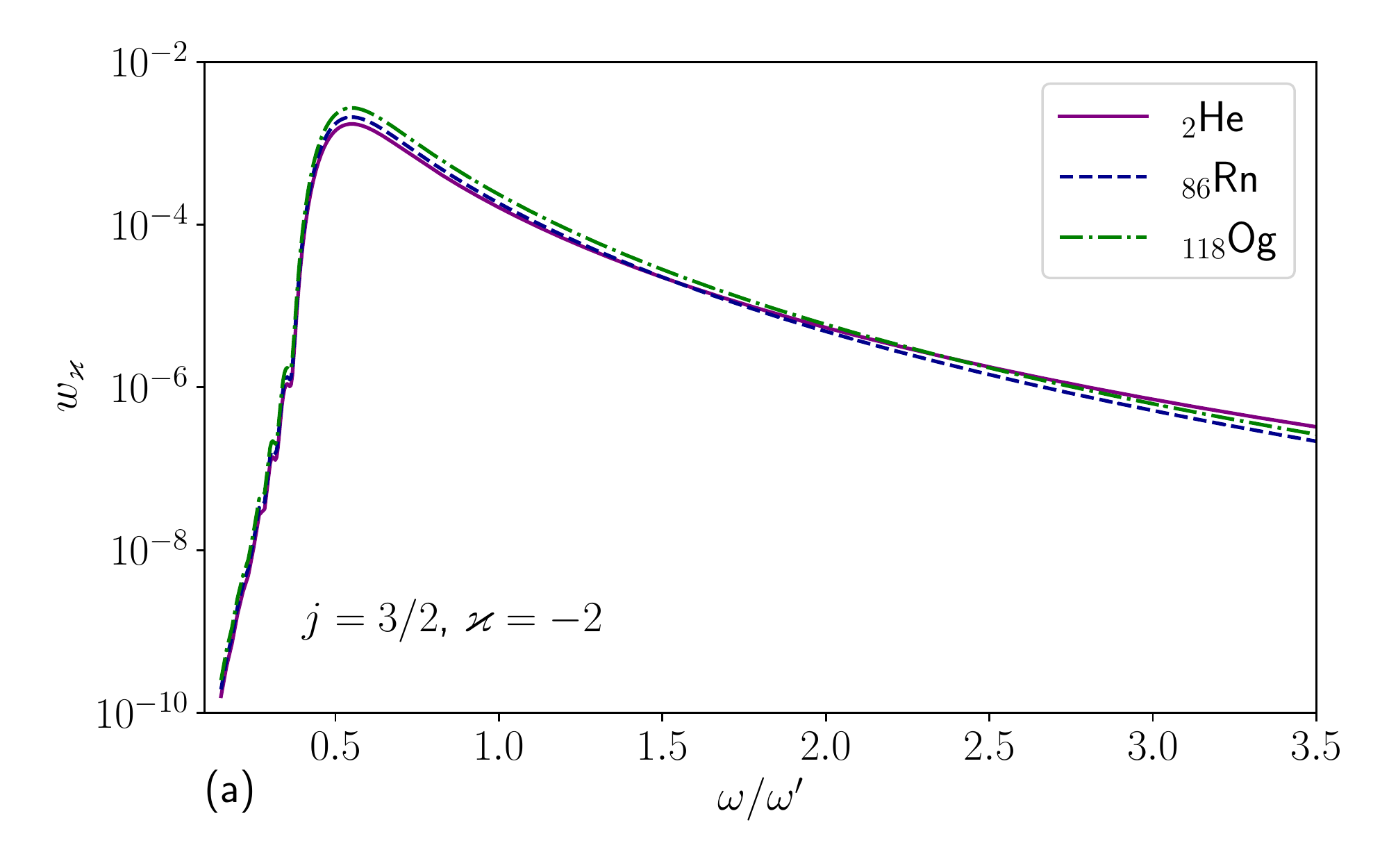}
	\includegraphics[width=0.45\textwidth]{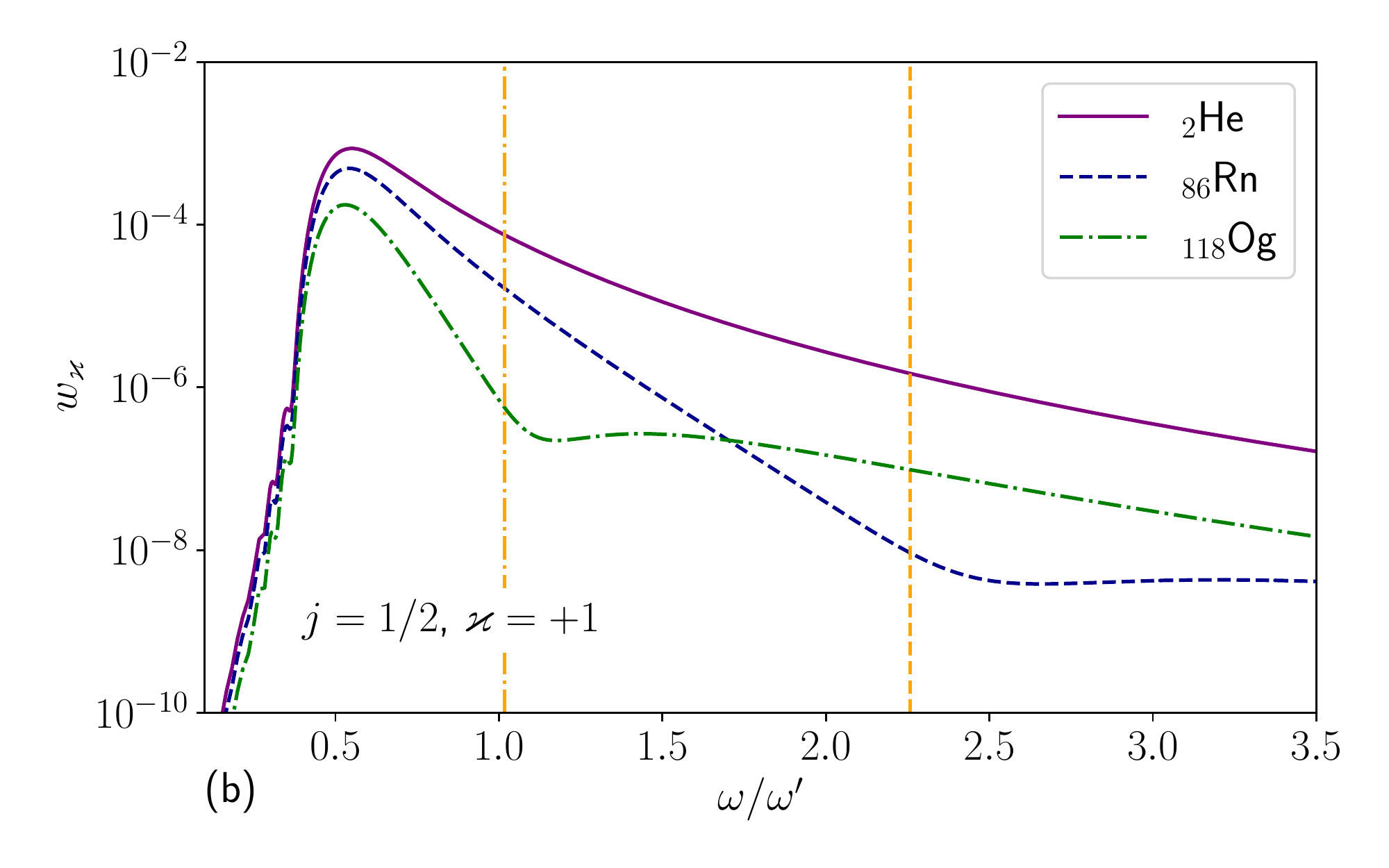}
	\caption{Probability of photoionization of hydrogen-like ions $_{2}$He$^{1+}$, $_{86}$Rn$^{85+}$ and $_{118}$Og$^{117+}$ as a function of laser frequency for two possible ionization channels. 
	Vertical lines indicate the laser frequency value $\omega=mc^2/\omega'$ for the cases of $_{118}$Og$^{117+}$ and $_{86}$Rn$^{85+}$ respectively.
	}
	\label{f:wdepkp}
\end{figure*}

In contrast to the nonrelativistic approach, Eq.~(\ref{dwke}) allows to extract information about two ionization channels (\ref{chan1}), (\ref{chan2})  with different values of the total angular momentum $j$ and the quantum number $\varkappa$.
The corresponding plots are shown in Fig.~\ref{f:wdepkp}. 

The probability of the  main channel with $\varkappa = -2$ may be scaled with a good accuracy and qualitatively reproduces the non relativistic result.
On the other hand, the ionization channel with $\varkappa = +1$ exhibits a local minimum.
This minimum is located near the frequency value of $\hbar\omega \approx mc^2$ in common units. 
This feature is absent in the channel with $\varkappa = -2$.

\begin{figure}
	\centering
	\includegraphics[width=0.45\textwidth]{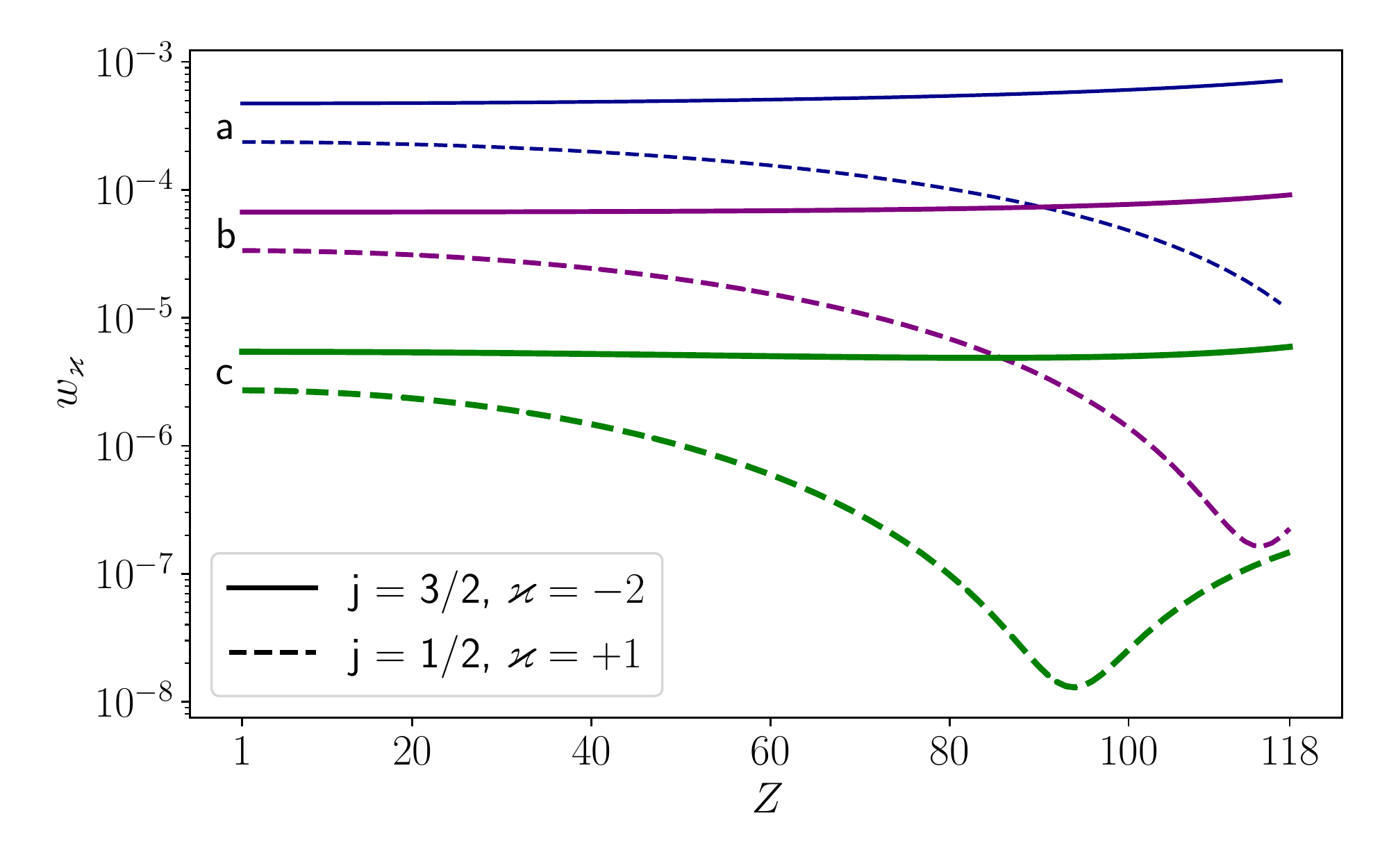}
	\caption{Probability of photoionization of heavy ions as a function of nucleus charge for various values of laser frequency,
	(a) $\omega = 0.8\omega'$,
	(b) $\omega = 1.2\omega'$,
	(c) $\omega = 2.0\omega'$.
	}
	\label{f:zdep}
\end{figure}

Suppression of the photoionization probability in the channel $\varkappa = +1$ also manifest itself clearly on the dependence of the photoionization probability on the nuclear charge $Z$ (Fig.~\ref{f:zdep}).

\begin{figure}
	\centering
	\includegraphics[width=0.45\textwidth]{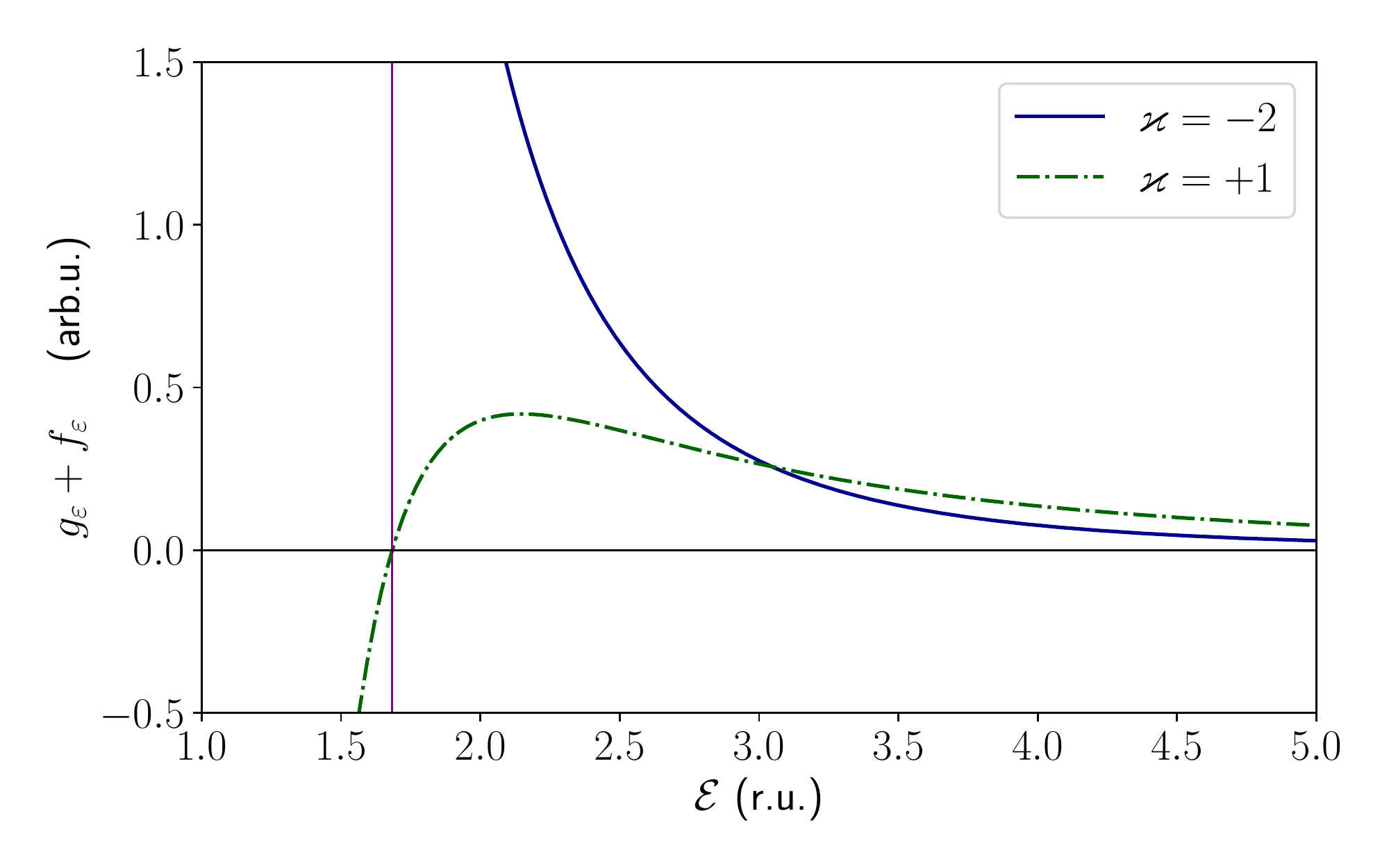}
	\caption{Dependence of the quantity $(g_\varepsilon + f_\varepsilon)$ on electron energy for $Z=86$.
	Vertical line is located at $\mathcal{E}^* = 1 + \mathcal{E}_{0}$
	}
	\label{f:gf}
\end{figure}

To clarify the origin of this suppression let us consider the factor 
$(g_\varepsilon + f_\varepsilon)$ entering Eq.~(\ref{dwke}).
If $\varkappa = -2$, the indicated factor is a monotonic positive function of the electron energy $\mathcal{E}$ and never equals to zero. 
On the other hand, when $\varkappa = +1$, this factor vanishes at energy value
\begin{equation}
\label{eo}
	\mathcal{E}^* = 1 + \mathcal{E}_{0}.
\end{equation}
This property is illustrated in Fig.~\ref{f:gf} for the case of ${Z = 86}$.

At the same time, in the perturbative case the most likely electronic transitions correspond to ionization into the energy range near 
\begin{equation}
\label{em}
	\mathcal{E}_m \approx \mathcal{E}_{0} + \omega.
\end{equation}
Thus, if $\omega \approx 1$ then $\mathcal{E}_m = \mathcal{E}^*$ and the factor 
$(g_\varepsilon + f_\varepsilon)$ and, hence, probability $dw_\varkappa(\mathcal{E})$ vanishes in the energy range giving the greatest contribution. 
This leads to probability suppression in the channel $\varkappa = +1$.

\begin{figure*}
	\centering
	\includegraphics[width=0.45\textwidth]{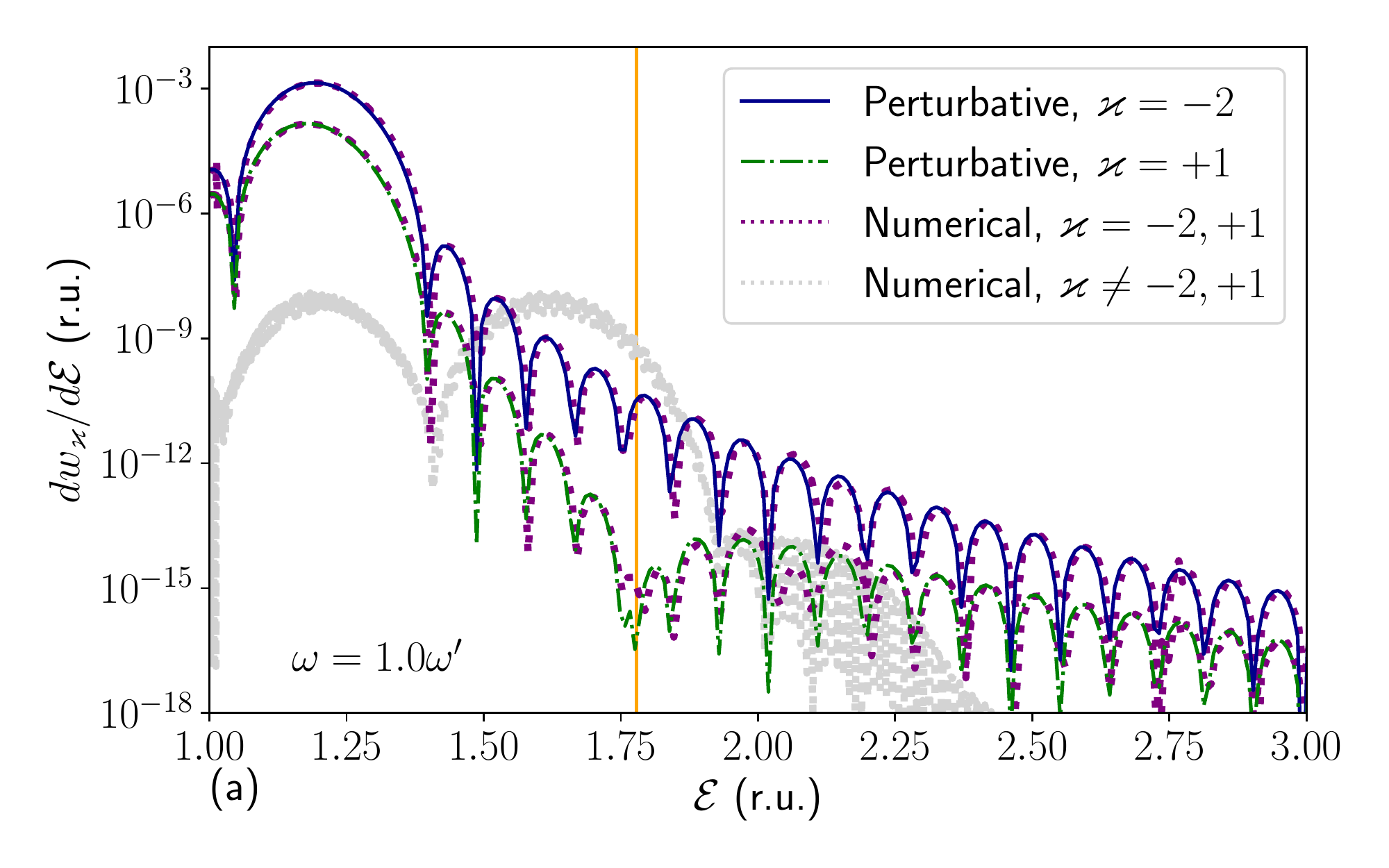}
	\includegraphics[width=0.45\textwidth]{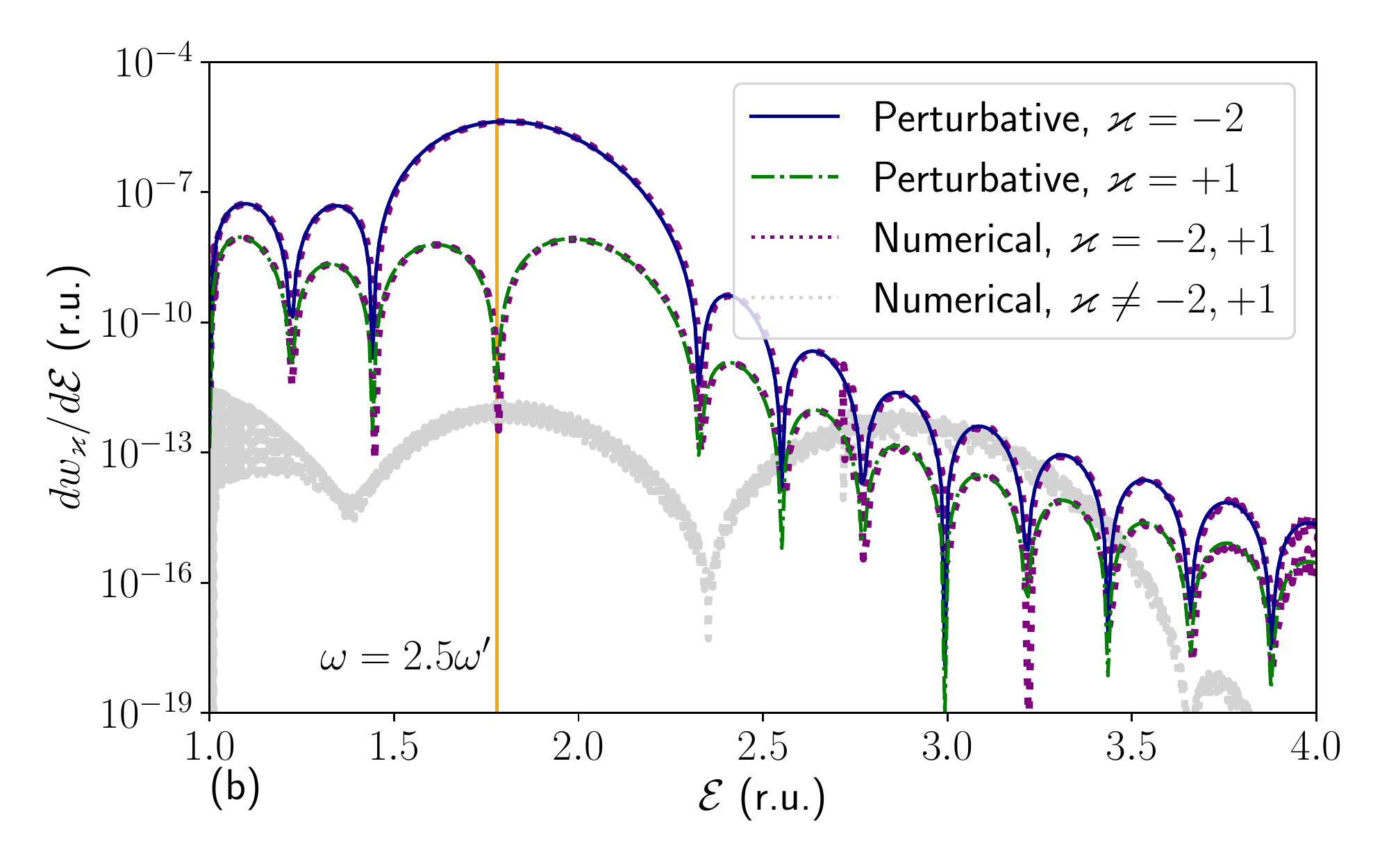}
	\caption{Electron spectra of photoionization of a $_{86}$Rn$^{85+}$ ion, obtained within perturbation theory and numerically.
	(a) Laser frequency $\omega = 1.0\omega'$; 
	(b) laser frequency $\omega = 2.5\omega'$ (or	$\omega \approx 1.1$).
	The vertical line is located at $\mathcal{E}^*=\mathcal{E}_0 + 1$.
 	}
	\label{f:espec}
\end{figure*}

Figure \ref{f:espec} shows energy spectra for ionization of $_{86}$Rn$^{85+}$ by a laser pulse with frequency
$\omega = 1.0\omega'$ ($\hbar\omega \approx 0.44mc^2$) and $\omega = 2.5\omega'$ ($\hbar\omega \approx 1.1mc^2$).
In the first case $\mathcal{E}^* > \mathcal{E}_m$ and suppression of the secondary channel with $\varkappa = +1$ is absent. 
In the second case the laser frequency corresponds to the location of minima at figs.~\ref{f:wdepkp} and \ref{f:zdep} when the condition $\mathcal{E}^* \approx \mathcal{E}_m$ is true.

A characteristic feature of the ionization spectra shown in Fig.~\ref{f:espec} is their oscillating background.
It originates from the integral over time in the process probability (\ref{dwke}),
\begin{equation}
    I  = \left|\int{E(t) e^{-i\omega_{fi}t} dt}\right|^2,
\end{equation}
where $\omega_{fi} = \mathcal{E} -\mathcal{E}_0$.
To clarify qualitatively this behavior, we note that the electric field of the laser pulse (\ref{Et}) is expressed via harmonic functions.
Therefore, the result of integration contains harmonic functions of combined frequencies, e.g. 
$I \propto \sin(\omega \pm \omega_{fi} \pm \nu_p)\tau$ etc, where $\nu_p$ is some frequency associated with the pulse envelope.
Apparently, oscillation period of these functions with respect to electron energy $\mathcal{E}$ is in inverse proportion to the pulse duration $\tau$.
Indeed, in both cases shown in Fig.~\ref{f:espec} the X-ray pulse has the same number of optical cycles.
As a result, the oscillations are wider in the case of higher frequency.

The perturbative probability given by Eq.~(\ref{dwke}) have been compared to numerically calculated ionization spectra.

As is seen, the feature in the secondary channel with $\varkappa = +1$ is confirmed by nonperturbative calculation.
We should note, however, that both the numerical and perturbative results are obtained within the dipole approximation with respect to the laser potential. 
Thus, the question if this feature is purely a dipole effect or it will be observed in the nondipole treatment as well needs some further investigation.

In addition to dipole transitions with spin-orbit quantum number values of $\varkappa = -2$ and $\varkappa=+1$, the nonperturbative numerical treatment allows us to calculate the probability of transitions that are forbidden in the perturbation theory.
In Fig.~\ref{f:espec}, the gray lines show the total probability density for the ionization channels with $\varkappa \neq -2$ and $\varkappa \neq +1$.
One can see, that their total contribution is negligible under the considered conditions.
However, their magnitude can be noticeable at high energy part of the ionization spectrum.

%%%%%%%%%%%%%%%%%%%%%%%%%%%%%%%%%%%%%%%%%%%%%%%%%%%%%%%%%%%%%%%%%%%%%%%%%%%%%%%
\section{Conclusions}
\label{sec:conclusions}
In the present work we develop relativistic perturbative approach to K-shell photoionization of a heavy ion by a short intense laser pulse. 
To describe the electron motion, we use solutions to the Dirac equation, while the laser potential is taken into account as a perturbation to the Coulomb hamiltonian in the dipole approximation.
Within this approach, two channels are allowed for ionization from the ground state to the continuum, that differ by the value of the total angular momentum $j$ and spin-orbit quantum number $\varkappa$. 
For these channels, the differential ionization probability is found in a simple analytical form.

The relativistic effects connected mostly with electron motion in a heavy ion. 
As a consequence, the relativistic scaling of the laser parameters allows to transform the total probability to the non-relativistic result with good accuracy.
At the same time, suppression of ionization with final electron having angular momentum $j = 1/2$ (and $\varkappa = +1$) is observed for photon energy of about $\hbar\omega \approx mc^2$.
This effect is purely relativistic and is confirmed by nonperturbative numerical calculation of ionization spectra.

\ack
This work is partially supported by the grant of the National Academy of Sciences of Ukraine (No. 0117U001760) under the Targeted research
program ``Collaboration in advanced international projects on high-energy physics and nuclear physics''

\end{document}